\documentclass[prd,superscriptaddress,amsfonts,amssymb,amsmath,showpacs,twocolumn]{revtex4-2}
\usepackage{bm}
\usepackage{amsfonts}
\usepackage{latexsym}
\usepackage{graphicx}
\usepackage{amsmath}
\usepackage{palatino}
\usepackage{mathpazo}
\usepackage{textcomp}
\usepackage{float}
\usepackage{booktabs}
\usepackage{dcolumn}
\usepackage{booktabs}
\usepackage{multirow}
\usepackage{hyperref}
\hypersetup{colorlinks,citecolor=blue}
\usepackage{amsmath}
\usepackage{xcolor}
\usepackage{orcidlink}
\usepackage[caption=false]{subfig}
\usepackage{commath}
\usepackage{tabularx}
\def\jnl@style{\it}
\def\aaref@jnl#1{{\jnl@style#1}}

\def\aaref@jnl#1{{\jnl@style#1}}

\def\aj{\aaref@jnl{AJ}}                   
\def\apj{\aaref@jnl{ApJ}}                 
\def\apjl{\aaref@jnl{ApJ}}                
\def\apjs{\aaref@jnl{ApJS}}               
\def\apss{\aaref@jnl{Ap\&SS}}             
\def\aap{\aaref@jnl{A\&A}}                
\def\aapr{\aaref@jnl{A\&A~Rev.}}          
\def\aaps{\aaref@jnl{A\&AS}}              
\def\mnras{\aaref@jnl{Mon.~Not.~Roy.~Astron.~Soc.}}             
\def\prd{\aaref@jnl{Phys.~Rev.~D}}        
\def\prc{\aaref@jnl{Phys.~Rev.~C}}  
\def\prl{\aaref@jnl{Phys.~Rev.~Lett.}}    
\def\qjras{\aaref@jnl{QJRAS}}             
\def\skytel{\aaref@jnl{S\&T}}             
\def\ssr{\aaref@jnl{Space~Sci.~Rev.}}     
\def\zap{\aaref@jnl{ZAp}}                 
\def\nat{\aaref@jnl{Nature}}              
\def\aplett{\aaref@jnl{Astrophys.~Lett.}} 
\def\apspr{\aaref@jnl{Astrophys.~Space~Phys.~Res.}} 
\def\physrep{\aaref@jnl{Phys.~Rep.}}      
\def\physscr{\aaref@jnl{Phys.~Scr}}       
\def\commat{\aaref@jnl{Comm.~Math.~Phys.}}              
\def\science{\aaref@jnl{Science}}               
\def\cqg{\aaref@jnl{Classical Quant.~Grav.}}            
\def\jpcs{\aaref@jnl{JPCS}}                                     
\def\ijmpd{\aaref@jnl{Int.~J.~Mod.~Phys.~D}}                    
\def\grg{\aaref@jnl{Gen.~Relat.~Gravit.}}               
\def\rpp{\aaref@jnl{Rep.~Prog.~Phys.}}          
\def\npa{\aaref@jnl{Nucl.~Phys.~A}}        
\def\lrr{\aaref@jnl{Living Rev.~Rel.}}                   
\def\jcap{\aaref@jnl{J.~Cosmology Astropart.~Phys.}}    
\def\rmp{\aaref@jnl{Rev.~Mod.~Phys.}}   
\def\epjc{\aaref@jnl{Eur.~Phys.~J.~C}}

\allowdisplaybreaks[1]

\begin{document}

\color{black}       

\title{Exploring Universe acceleration through observational constraints via Hubble parameter reconstruction}

\author{M. Koussour\orcidlink{0000-0002-4188-0572}}
\email[Email: ]{pr.mouhssine@gmail.com}
\affiliation{Department of Physics, University of Hassan II Casablanca, Morocco.} 

\author{N. Myrzakulov\orcidlink{0000-0001-8691-9939}}
\email[Email: ]{nmyrzakulov@gmail.com}
\affiliation{L. N. Gumilyov Eurasian National University, Astana 010008,
Kazakhstan.}

\author{M. K. M. Ali\orcidlink{0000-0000-0000-0000}}
\email[Email: ]{mkali@imamu.edu.sa}
\affiliation{Department of Physics, College of Science, Imam Mohammad Ibn Saud Islamic University (IMSIU),\\
Riyadh 13318, Saudi Arabia.}

\date{\today}

\begin{abstract}
In this article, we introduce an innovative parametric representation of the Hubble parameter, providing a model-independent means to explore the dynamics of an accelerating cosmos. The model's parameters are rigorously constrained through a Markov Chain Monte Carlo (MCMC) approach, leveraging a comprehensive dataset consisting of 31 data points from cosmic chronometers (CC), 1701 updated observations of Pantheon supernovae type Ia (SNeIa), and 6 data points from baryonic acoustic oscillations (BAO). Our analysis delves into the behavior of various cosmological parameters within the model, including the transition from a decelerating phase to an accelerating one, as well as the density parameters and the equation of state (EoS) parameter. The outcomes of our investigation reveal that the equation of state parameter aligns with characteristics reminiscent of the phantom model, supporting the prevailing understanding of our universe's current state of acceleration. This research contributes valuable insights into the ongoing cosmic expansion and underscores the utility of our novel parametric approach.
\end{abstract}

\maketitle

\section{Introduction}
\label{sec1}

Over the past two decades, a compelling body of evidence has emerged from various astronomical observations and experiments, reshaping our understanding of the universe's fate. Type Ia Supernovae (SNe) searches \cite{Riess, Perlmutter}, Cosmic Microwave Background Radiation (CMBR) studies \cite{R.R., Z.Y.}, the Wilkinson Microwave Anisotropy Probe (WMAP) experiment \cite{C.L.,D.N.}, and Baryon Acoustic Oscillation (BAO) measurements \cite{D.J., W.J.} have collectively and convincingly pointed to a remarkable phenomenon: the accelerating expansion of our cosmos. This revelation has ignited profound inquiries into the ultimate destiny of our universe.

The driving force behind this cosmic expansion is attributed to a mysterious entity known as Dark Energy (DE) \cite{Peebles,Padmanabhan}. DE is characterized by an Equation of State (EoS) parameter denoted as $\omega_{DE}$, which represents the ratio of spatially homogeneous pressure ($p_{DE}$) to the energy density ($\rho_{DE}$) of DE. Understanding the fundamental nature of DE holds the key to deciphering the eventual outcome of our universe. However, unraveling the enigma of DE has proven to be a formidable challenge. Recent cosmological observations have revealed significant uncertainties in pinning down the precise nature of DE, as indicated by the range of possible values for the EoS parameter. These values encompass $\omega_{DE}<-1$, $\omega_{DE}=-1$, and $\omega_{DE}>-1$, introducing a degree of ambiguity into our cosmological understanding. For instance, data combined from Hubble constant measurements, SNe observations, CMB data, and BAO measurements, as reported by WMAP \cite{G.H.}, suggest a value of $\omega_{DE}=-1.084\pm0.063$. In 2015, the Planck collaboration reported a slightly different value, $\omega_{DE}=-1.006\pm0.0451$ \cite{P15}. Subsequently, in 2018, their findings indicated $\omega_{DE}=-1.028\pm0.032$ \cite{N18}. These variations underscore the complexity of the DE conundrum and highlight the need for ongoing research and precision measurements to elucidate the true nature of this enigmatic cosmic component. The fate of our universe is intricately tied to our ability to unravel the mysteries of DE, making it a topic of paramount importance in modern cosmology.

Within the realm of cosmology, the cosmological constant denoted as $\Lambda$ in the framework of General Relativity (GR) stands as the most straightforward explanation for DE, characterized by an EoS parameter $\omega_{DE}$ equal to $-1$. However, a significant quandary emerges when we compare the observed value of the cosmological constant $\Lambda$ with its anticipated value stemming from quantum gravity theories \cite{S.W.}. This glaring discrepancy between observed and expected values is what cosmologists refer to as the "cosmological constant problem" a conundrum that continues to perplex researchers. In the quest to understand the dynamic nature of DE, various time-varying models have garnered attention. Among these, the quintessence DE model, characterized by an EoS parameter within the range $-1<\omega_{DE}<-\frac{1}{3}$, has gained prominence. In quintessence models, the density of DE diminishes as the universe evolves \cite{RP,M.T.,LX}. While this framework provides a viable alternative to the cosmological constant, it raises important questions about the nature and evolution of DE. On the intriguing end of the spectrum, we find phantom energy, a form of DE characterized by $\omega_{DE}<-1$. This unconventional model has piqued the interest of theorists due to its peculiar properties. In the phantom energy scenario, DE exhibits unbounded growth, resulting in an extreme future expansion that culminates in a finite-time future singularity. This unusual behavior challenges our understanding of cosmic evolution and has prompted investigations into the classification of singularities \cite{KI,DB,DB-2}. Beyond these paradigms, researchers have explored alternative approaches to understanding DE, notably through modified theories of gravity \cite{L.A.,SA,R.F.}. These modified gravity theories offer alternative explanations for the observed cosmic acceleration, diverging from the conventional framework of $\Lambda$CDM cosmology. 

Among the various approaches to understanding the complex dynamics of the cosmos, the reconstruction technique stands out as a promising avenue. This method employs rigorous statistical treatments to craft a robust kinematic model, offering a compelling means to elucidate the cosmic narrative. What makes the reconstruction technique particularly intriguing is its recent surge in popularity, driven by its remarkable ability to align with observational data, rendering it independent of the underlying gravity model. Reconstruction methods can be broadly categorized into two types: parametric and non-parametric reconstruction \cite{Mukherjee1}. In the case of non-parametric reconstruction, models are directly derived from observational data through meticulous statistical procedures. On the other hand, parametric reconstruction initially formulates a kinematic model with free parameters, subsequently narrowing down the parameter space through a rigorous statistical analysis of observational data. This approach represents a conceptually straightforward means of addressing some of the longstanding shortcomings of the standard cosmological model, including late-time acceleration, the enigmatic cosmological constant problem, and the enigma of the universe's initial singularity. In recent research, there has been a concerted effort to employ parametric reconstruction techniques to shed light on cosmic acceleration. For instance, in a notable study \cite{Mukherjee2}, researchers investigated the accelerating universe by parametrically reconstructing the jerk parameter, a quantity that characterizes changes in acceleration. This approach offers valuable insights into the evolution of the Hubble parameter, a fundamental cosmic parameter. Similarly, another group of scientists, as seen in the work of Campo et al. \cite{Del}, focused on parametrizing the deceleration parameter. This parametrization technique proved to be successful in predicting thermal equilibrium, a crucial aspect of the universe's thermal history. Furthermore, researchers like Gong et al. \cite{Gong} have explored parametrization techniques to better understand the EoS for DE, providing valuable tools for probing the enigmatic nature of this cosmic component.

In the extensive landscape of cosmological research, a plethora of studies have delved into the parametrization of key cosmological parameters such as the deceleration parameter, jerk parameter, and the EoS parameter \cite{DP1,DP2,DP3,CPL1,CPL2,JBP,MZ,Mamon,Hernandez,Koussour1,Koussour2,Koussour3,Koussour4}. These investigations have contributed significantly to our understanding of the evolving universe and its intriguing dynamics. However, it is noteworthy that relatively fewer studies have ventured into the parametrization of one of the most foundational and fundamental cosmological quantities—the Hubble parameter. The Hubble parameter stands as a cornerstone in cosmology, serving as a critical indicator of the universe's expansion and evolution. Given its paramount importance, there exists a compelling motivation to explore and develop parametric forms of the Hubble parameter \cite{H1,H2}. In this context, our research endeavors have led to the construction of a novel parametric representation of the Hubble parameter, with a primary focus on constraining the Hubble constant. The central objective of this method is to craft a viable model that accurately captures the dynamics of an accelerating universe. An inherent strength of this approach lies in its independence from any specific gravity model, which grants it a degree of versatility and applicability to a wide range of cosmological scenarios. To validate this newly developed parametric form of the Hubble parameter, we rigorously examine the solutions of Einstein's field equations, particularly in the context of an isotropic and homogeneous universe. 

The structure of this paper is outlined as follows: Sec. \ref{sec2} furnishes a comprehensive examination of the FLRW universe and the underlying cosmological model. In Sec. \ref{sec3}, we introduce a novel parametric representation for the Hubble parameter, offering an innovative perspective on cosmic dynamics. Sec. \ref{sec4} is dedicated to a detailed exploration of our approach to observational data and research methodology. In Sec. \ref{sec5}, we present the compelling outcomes derived from our data analysis, highlighting the key cosmological parameters. Finally, Sec. \ref{sec6} serves as a summary and synthesis of our results and discussions, encapsulating the essence of our research.

\section{Cosmological model}
\label{sec2}

In the present paper, we consider a spatially flat, homogeneous, and isotropic Friedmann-Lemaître-Robertson-Walker (FLRW) Universe described by the metric \cite{Ryden} as
\begin{equation}
ds^{2}=dt^{2}-a^{2}(t)[dr^{2}+r^{2}(d{\theta }^{2}+sin^{2}\theta d{\phi }%
^{2})],  \label{FLRW}
\end{equation}%

In this context, we have a Universe characterized by a scale factor $a(t)$ that varies with time. Within this Universe, there exist two distinct perfect fluids. The first fluid corresponds to ordinary matter and has negligible pressure. The second fluid represents DE. Under the assumption of natural units ($8\pi G=c=1$), we can express Einstein's field equations as follows,
\begin{eqnarray}
\label{F1}
  3H^2&=&\rho_{M} +\rho_{DE},\\
  2\dot{H} +3H^2&=& -p_{DE},\label{F2}
\end{eqnarray} 
where $H=\frac{\dot{a}}{a}$ plays a crucial role as it quantifies the rate of cosmic expansion in this context. Also, $\rho_{M}$, $\rho_{DE}$, and $p_{DE}$ represent the energy densities of ordinary matter, the DE, and the pressure associated with the DE, respectively.

Moreover, the energy density of ordinary matter undergoes evolution with the scale factor given by
\begin{equation}
\label{matter}
\rho_{m}=\rho_{m0}a^{-3}=\rho_{m0}(1+z)^{3},
\end{equation}
where $\rho_{m0}$ represents the current value of matter-energy density. Here, $z$ denotes the cosmological redshift defined as $z=\frac{1}{a(t)}-1$. Thus, it is possible to express the derivatives with respect to cosmic time in terms of derivatives with respect to redshift using the following relation: $\frac{d}{dt}=-\left(
1+z\right) H\left( z\right) \frac{d }{dz}$. The derivative of the Hubble parameter with respect to cosmic time can be expressed as a function of the redshift as
\begin{equation}
\overset{.}{H}=-\left(
1+z\right) H\left( z\right) \frac{dH\left( z\right) }{dz}.
\label{t_z}
\end{equation}

In order to assess the nature of cosmological expansion, whether it is accelerating or decelerating, we introduce the deceleration parameter $q$, given by the expression:
\begin{equation}\label{q}
q(z)=-1-\frac{\dot{H}}{H^2}=\frac{(1+z)}{H(z)}\frac{dH(z)}{dz}-1.
\end{equation}

A positive value of $q$ indicates a deceleration in the expansion of the Universe. At $q=0$, the expansion of the Universe proceeds at a constant rate. On the other hand, if $-1 <q < 0$, it signifies accelerating growth of the Universe. Particularly, when $q = -1$, the Universe undergoes exponential expansion, commonly known as de Sitter expansion. Furthermore, for $q < -1$, the Universe exhibits super-exponential expansion.

The density parameters $\Omega_{M}$, and $\Omega_{DE}$ are significant cosmological parameters that provide crucial information about the matter composition of the Universe. They are defined as follows:
\begin{equation}\label{OmegaM}
 \Omega_M(z)=\frac{\rho_{M}}{3H^2},
\end{equation}
\begin{equation}\label{OmegaDE}
\Omega_{DE}(z)=1-\Omega_{M}(z),
\end{equation}

In order to deepen our comprehension of the accelerated period, we introduce the equation of state (EoS) parameter for DE $\omega_{DE}(z)$, which is given by
\begin{equation}
\omega_{DE}(z)=\frac{p_{DE}}{\rho_{DE} }=\frac{-1-\frac{2 \dot{H}}{3 H^2}}{\Omega_{DE}}.
\end{equation}

Hence, this yields
\begin{equation}\label{wDE}
\omega_{DE}(z)=\frac{2(1+z)H\frac{dH}{dz}-3H^2}{3H^2-\rho_{M0}(1+z)^3},
\end{equation}

By using Eqs. (\ref{F1}) and (\ref{F2}), we can deduce the equation that describes the acceleration \cite{Sahni1},
\begin{equation}
    \frac{\overset{..}{a}}{a}=-\frac{1}{6}\left( \rho_{M} +\rho_{DE}+3p_{DE}\right).
\end{equation}

Hence, according to the derived expression, the current model foresees acceleration ($\overset{..}{a}>0$) only when $\omega<-\frac{1}{3}$. The EoS parameter serves not only to differentiate between the universe's decelerated and accelerated expansion phases but also to categorize them. The epoch of matter domination corresponds to a value of $\omega$ equal to zero. In the current epoch of accelerated expansion, $\omega_{DE}< -1$ pertains to the phantom era, while $-1/3 < \omega_{DE}< -1$ represents the quintessence phase. For $\omega_{DE} = -1$, this value represents the cosmological constant in the $\Lambda$CDM model, associated with DE.

\section{The parametrization}
\label{sec3}

Eqs. (\ref{F1})-(\ref{matter}) consist of three sets of equations with four unknown parameters: $\rho_{M}$, $\rho_{DE}$, $p_{DE}$ (or $\omega_{DE}$), and $H$. To find a solution, it is necessary to make an ansatz. In cosmology, the Hubble parameter $H(z)$ plays a fundamental role in characterizing the expansion rate of the Universe. To model and understand the behavior of the Hubble parameter as a function of redshift $z$, a common approach is to adopt a parametric form. This parametrization involves introducing a set of parameters and a functional form that represents the Hubble parameter's evolution with redshift. Various parametric forms have been proposed, each capturing different cosmological scenarios and behaviors. Recently, Koussour et al. \cite{Koussour1} proposed a comprehensive parameterization approach of the dimensionless Hubble parameter within the framework of scalar field DE models, which is represented by the equation: $E^2(z)=\frac{H^2(z)}{H_0^2}=A(z)+\beta (1+\gamma B(z))$, where $\beta$, $\gamma$ are free parameters, and $A(z)$, $B(z)$ are functions of the redshift $z$. The authors incorporate correction terms associated with DE in the context of the $\Lambda$CDM model, expressed as $A(z)=\alpha(1+z)^3$ and $B(z)=\frac{z}{1+z}$, where $\alpha=\Omega_{m0}$. In this article, we adopt this approach and propose a parametrization of the Hubble parameter as an explicit function of cosmic redshift in the following form:
\begin{equation}
\label{Hz}
    H(z)=H_{0} \sqrt{\alpha  (1+z)^3+\beta  \left(1+\gamma\frac{ z(1+z)}{1+z^2}\right)},
\end{equation}
where $\alpha$, $\beta$, and $\gamma$ are free parameters, and $H_{0}$ represents the present value of the Hubble parameter. To recover the $\Lambda$CDM model using this parametrization, appropriate values for $\alpha$, $\beta$, and $\gamma$ must be set. The $\Lambda$CDM model corresponds to a specific case within this parametrization, with the values chosen as follows: $\alpha=\Omega_{M0}$, $\beta=\Omega_{\Lambda}$, and $\gamma=0$. Notably, $H(z=0)=H_0$, imposes an additional constraint on the model's parameters, reducing their number and leading to the relationship $\alpha+\beta=1$. In the context of the $\Lambda$CDM model, the evolution of the Hubble parameter is described by
\begin{equation}
\label{HLCDM}
    H(z)=H_{0} \sqrt{\Omega_{M0}  (1+z)^3+\Omega_{\Lambda} },
\end{equation}
where $\Omega_{M0}$ and $\Omega_{\Lambda}$ represent the present density parameters of matter and DE (interpreted as a cosmological constant), respectively. These density parameters are subject to a significant constraint, namely $\Omega_{M0}+\Omega_{\Lambda}=1$.

The evolution of the deceleration parameter with redshift is derived as
\begin{equation}
    q(z)=\frac{3 \Omega_{M0} (1+z)^3}{2 \left[(1-\Omega_{M0})+\Omega_{M0} (1+z)^3\right]}-1.
\end{equation}

Here, we consider the numerical values $H_0=(67.4 \pm 0.5)$ $km/s/Mpc$ and $\Omega_{M0}=0.315 \pm 0.007$ obtained from the latest Planck data \cite{N18}. Consequently, the present-day value of the deceleration parameter is $q_0=-0.528$. As for the evolution of the matter density parameter concerning the redshift, we obtain the expression $\Omega_{M}(z)=0.315 (1+z)^3$.

Given the particular choice of 
$H(z)$ as provided in Eq. (\ref{Hz}), Eqs. (\ref{q}), (\ref{OmegaM}), (\ref{OmegaDE}), and (\ref{wDE}) can be reformulated as follows:
\begin{equation}\label{q1}
q(z)=\frac{(1+z) \left(3 \alpha  (1+z)^2+\gamma(\alpha -1)\frac{(z-2) z-1}{\left(1+z^2\right)^2}\right)}{2 \left(\alpha  (1+z)^3+(1-\alpha ) \left(1+\gamma\frac{z (1+z)}{1+z^2}\right)\right)}-1,
 \end{equation}
\begin{equation} \label{omegam1}
\Omega_{M}(z)=\frac{\Omega_{M0} (1+z)^3}{\alpha  (1+z)^3+(1-\alpha ) \left(1+\gamma\frac{ z (1+z)}{1+z^2}\right)},
\end{equation}
   \begin{equation}
  \Omega_{DE}(z)=1-\frac{\Omega_{M0} (1+z)^3}{\alpha  (1+z)^3+(1-\alpha ) \left(1+\gamma\frac{ z (1+z)}{1+z^2}\right)},
\end{equation}
and
\begin{widetext}
  \begin{equation} \label{wphi1}
 \omega_{DE}(z) =\frac{ (1+z) \left(3 \alpha  (1+z)^2+\gamma(\alpha -1)\frac{(z-2) z-1}{\left(1+z^2\right)^2}\right)-3 \left(\alpha  (1+z)^3+(1-\alpha ) \left(1+\gamma\frac{z (1+z)}{1+z^2}\right)\right)}{3 \left(\alpha  (1+z)^3+(1-\alpha ) \left(1+\gamma\frac{z (1+z)}{1+z^2}\right)\right)-3\Omega_{M0}(1+z)^3}.
\end{equation}
\end{widetext}

\section{Examination of Observational Data and Research Approach}
\label{sec4}

For investigating the observational characteristics of our cosmological model, we utilize the latest cosmic Hubble $Hz$, SNe observations, and BAO. This involves 31 data points from the $Hz$ datasets, 1701 data points from the Pantheon+ SNe samples, and 6 data points from the BAO datasets. To analyze the data, we employ the Bayesian analysis and likelihood function, and to efficiently explore the parameter space, we employ the MCMC method using the emcee Python library \cite{Mackey}.

\subsection{$Hz$ dataset (CC)}
Using the CC (Cosmic Chronometers) method, the Hubble rate is determined by studying the oldest and most passively evolving galaxies, which are separated by a small redshift interval, employing the differential aging method \cite{D1,D2,D3,D4}. The method relies on the definition of the Hubble rate given by:
\begin{equation}
    H(z) = -\frac{1}{1+z} \frac{dz}{dt}.
\end{equation}

One of the notable advantages of the CC method is its capability to measure the Hubble parameter independently, without relying on any specific cosmological assumptions. As a result, it serves as a valuable tool to rigorously test and validate various cosmological models. The data points collected from Refs. \cite{Moresco/2018} were obtained using the CC method, encompassing a wide range of redshifts from 0.1 to 2. For the MCMC analysis, the chi-square function for the CC is expressed as 
\begin{equation}
\chi^2_{Hz}(H_{0},\alpha,\gamma)=\sum_{k=1}^{31}\left[\frac{(H_{th}(z_k,H_{0},\alpha,\gamma)-H_{obs}(z_k))^2)}{\sigma_H^2(z_k)}\right],
\end{equation}
where $H_{th}$ represents the theoretical estimation of the Hubble parameter for a specific model, which is characterized by the model parameters $H_{0}$, $\alpha$, and $\gamma$. On the other hand, $H_{obs}$ refers to the observed values of the Hubble parameter, while $\sigma_H$ denotes the associated error in the estimation.

\subsection{SNe dataset (Pantheon+)}

The Pantheon+ analysis represents an advancement over the original Pantheon analysis by incorporating an expanded dataset of supernovae (SNe), including those associated with galaxies for which Cepheid distances have been measured. This extensive dataset encompasses 1701 light curves from a total of 1550 SNe, spanning a redshift range from $0.001$ to $2.2613$. Data for this comprehensive compilation have been sourced from 18 separate studies \cite{pan1,pan2,pan3,pan4}. Of the 1701 light curves in this dataset, 77 are linked to galaxies known to contain Cepheid variables. The Pantheon+ compilation, when compared to the original Pantheon compilation by Scolnic et al. \cite{Scolnic}, offers notable improvements and enhancements. Most significantly, Pantheon+ boasts a larger sample size, particularly notable for its increased number of SNe at redshifts below 0.01. Furthermore, substantial efforts have been made to address and mitigate systematic uncertainties pertaining to redshift measurements, models for intrinsic scatter, photometric calibration, and peculiar velocities of SNe. It's important to note that, due to specific selection criteria, not all SNe from the original Pantheon compilation have been incorporated into the upgraded Pantheon+ compilation. Pantheon+ offers an added benefit by allowing for the simultaneous determination of the Hubble constant ($H_0$) alongside the model parameters. To achieve the optimal values for the independent parameters, it becomes imperative to optimize the $\chi^2$ function, as outlined below:
\begin{equation}
    \chi^2_{SNe}= \Delta\mu^T (C_{Sys+Stat}^{-1})\Delta\mu.
\end{equation}

Here, $C_{Sys+Stat}$ represents the covariance matrix for the Pantheon+ dataset, which incorporates a combination of systematic and statistical uncertainties. The term $\Delta\mu$ represents the residual distance and is defined as $\Delta\mu_k=\mu_k-\mu_{th}(z_k)$, where $\mu_k$ denotes the distance modulus of the $k^{th}$ SNe. It's important to emphasize that $\mu_k$ is computed as $\mu_k = m_{Bk} - M$, where $m_{Bk}$ corresponds to the apparent magnitude of the $k^{th}$ SNe and $M$ represents the reference magnitude for a SNe. 
The theoretical distance modulus $\mu_{th}$ is determined using the following formula:
\begin{equation}\label{Eq:mu}
    \mu_{th}(z,H_{0},\alpha,\gamma)=5\log_{10}\left(\frac{d_L(z,H_{0},\alpha,\gamma)}{1\text{ Mpc}} \right)+25,
\end{equation}

Here, $d_L$ denotes the model-dependent luminosity distance in megaparsecs (Mpc), and it is defined as:
\begin{equation}
d_L(z,H_{0},\alpha,\gamma)=\frac{c(1+z)}{H_0}\int_0^z \frac{dy}{E(y)},
\end{equation}
where the symbol $c$ stands for the speed of light. Additionally, it's worth noting that the parameters $M$ and $H_0$ exhibit degeneracy, especially when analyzing SNe data. However, taking into account the recent SH0ES results alleviates these limitations. Consequently, we can express the distance residual as 
\begin{equation}
    \Delta\Bar{\mu}=\begin{cases}
            \mu_k-\mu_k^{cd}, & \text{If $k$ is in Cepheid hosts}\\
            \mu_k-\mu_{th}(z_k), & \text{Otherwise}
         \end{cases},
\end{equation}

In this equation, $\mu_k^{cd}$ represents the Cepheid host-galaxy distance provided by SH0ES. When computing the covariance matrix for the Cepheid host-galaxy data, it can be merged with the covariance matrix for SNe. The resulting combined covariance matrix, denoted as $C^{SNe}_{Sys+Stat}+C^{cd}_{Sys+Stat}$, encompasses both the statistical and systematic uncertainties from the Pantheon+ dataset and the Cepheid host-galaxy data.

Consequently, the $\chi^2$ function employed for constraining cosmological models in the analysis is expressed as
\begin{equation}\label{eq:chiSNe}
    \chi^2_{SNe+}= \Delta\Bar{\mu} (C^{SNe}_{Sys+Stat}+C^{cd}_{Sys+Stat})^{-1}\Delta\Bar{\mu}^T.
\end{equation}

\subsection{$Hz+SNe+BAO$ dataset (Joint Analysis)}

Expanding upon our previous dataset, our analysis will incorporate additional BAO data sources. BAOs arise from density fluctuations in baryonic matter throughout the Universe. They originate from acoustic density waves in the early stages of the primordial plasma. These oscillations provide valuable insights because they can be utilized to derive important cosmological parameters related to DE. In this investigation, we incorporate the BAO dataset, which has been compiled from various surveys, including the 6dFGS, the SDSS, and the LOWZ samples of the BOSS \cite{BAO1, BAO2, BAO3, BAO4, BAO5, BAO6}. These surveys have produced highly precise measurements of the positions of BAO peaks in galaxy clustering across a range of redshifts. The characteristic scale of BAO, denoted as the sound horizon $r_s$ at the epoch of photon decoupling with a redshift of $z_{dec}$, is linked by the following equation:
\begin{equation}
r_{s}(z_{\ast })=\frac{c}{\sqrt{3}}\int_{0}^{\frac{1}{1+z_{\ast }}}\frac{da}{
a^{2}H(a)\sqrt{1+(3\Omega _{b,0}/4\Omega _{\gamma,0})a}},
\end{equation}

Here, the symbols $\Omega {b,0}$ and $\Omega {\gamma,0}$ denote the present-day density parameters for baryons and photons, respectively. The BAO dataset utilized in this study consists of six data points corresponding to the ratio  $d_{A}(z_{\ast })/D_{V}(z_{BAO})$. These specific data points have been sourced from Refs. cited in \cite{BAO1, BAO2, BAO3, BAO4, BAO5, BAO6}. In this context, the term $z_{\ast }\approx 1091$ represents the redshift value associated with photon decoupling, and $d_{A}(z_{\ast })=c\int_{0}^{z}\frac{dz'}{H(z')}$ signifies the comoving angular diameter distance at the time of decoupling. Additionally, we define the dilation scale as $D_{V}(z)=\left[czd_{A}^{2}(z)/H(z)\right] ^{1/3}$. The chi-square function, as outlined in \cite{BAO6}, is employed for the assessment of the Baryon Acoustic Oscillation (BAO) dataset. It can be formulated as 
\begin{equation}\label{4e}
\chi _{BAO}^{2}=X^{T}C_{BAO}^{-1}X,
\end{equation}
where 
\begin{equation}
X=\left( 
\begin{array}{c}
\frac{d_{A}(z_{\star })}{D_{V}(0.106)}-30.95 \\ 
\frac{d_{A}(z_{\star })}{D_{V}(0.2)}-17.55 \\ 
\frac{d_{A}(z_{\star })}{D_{V}(0.35)}-10.11 \\ 
\frac{d_{A}(z_{\star })}{D_{V}(0.44)}-8.44 \\ 
\frac{d_{A}(z_{\star })}{D_{V}(0.6)}-6.69 \\ 
\frac{d_{A}(z_{\star })}{D_{V}(0.73)}-5.45%
\end{array}%
\right) \,,
\end{equation}%
and the term $C_{BAO}^{-1}$ corresponds to the inverse of the covariance matrix \cite{BAO6}.

In our final step, we examine the combination of the aforementioned observational datasets. For our study, we will utilize the following combination:
\begin{eqnarray}
 \chi _{Joint}^{2}=\chi _{Hz}^{2}+\chi _{SNe}^{2}+\chi _{BAO}^{2}.
\end{eqnarray} 

The model parameters undergo constraint through the minimization of their respective $\chi^2$ values, which are intricately linked to the likelihood through $\mathcal{L} \propto \exp\left( -\frac{\chi^2}{2} \right)$. This process is achieved by employing the MCMC sampling technique in conjunction with the {\tt emcee} library.

\section{Outcomes of Data Analysis}
\label{sec5}
In the previous section, we delved into the intricacies of harnessing observational data and applying statistical tools to finely tune our model parameters, specifically $H_0$, $\alpha$, and $\gamma$. These parameters play a pivotal role in our quest to understand and characterize the universe's fundamental properties. Now, let's transition into the fascinating realm of observational results obtained through rigorous statistical analysis. Our efforts have yielded the following model parameter values:

For the $Hz$ dataset:
\begin{itemize}
    \item $H_0=67.8_{-1.8}^{+1.7}$,
    \item $\alpha=0.34\pm0.13$,
    \item $\gamma=-0.2_{-1.3}^{+1.2}$.
\end{itemize}

\begin{figure}[h]
\includegraphics[width=1.0\linewidth]{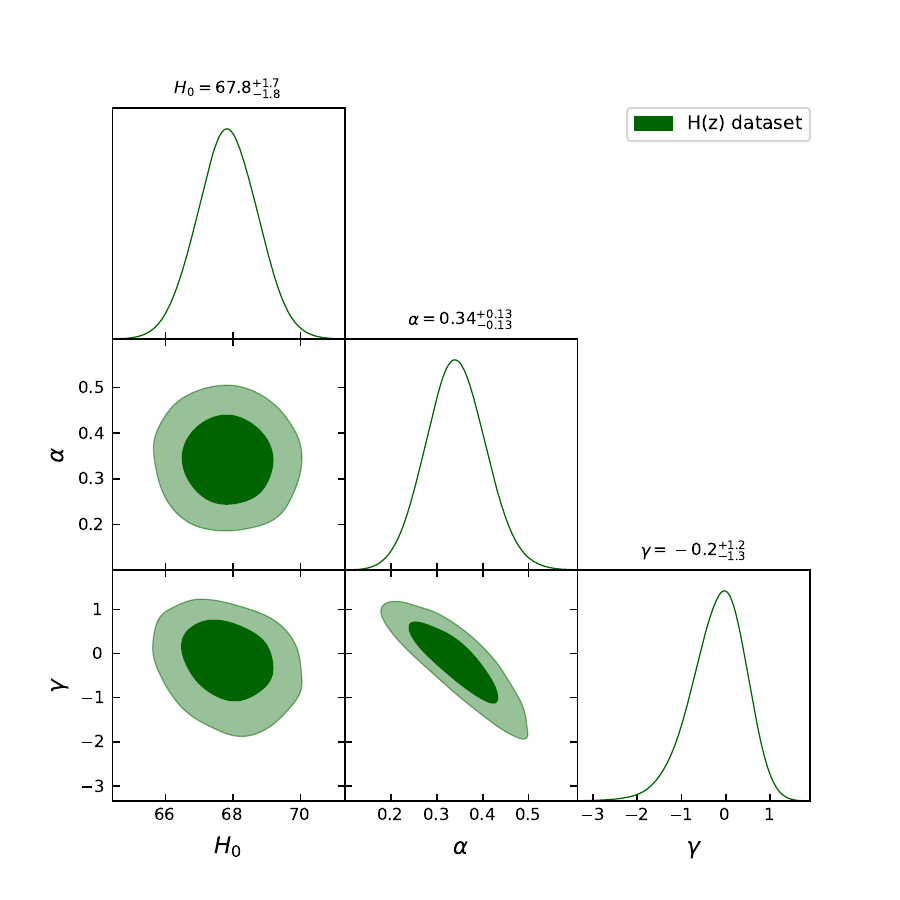}
     \caption{Contours of likelihood for $Hz$ datasets, indicating $1-\sigma$ and $2-\sigma$ confidence regions.}\label{F_Hz}
\end{figure}

For the newly released 1701 $SNe$ dataset:
\begin{itemize}
    \item $H_0=69.0_{-3.2}^{+3.3}$,
    \item $\alpha=0.49_{-0.31}^{+0.29}$,
    \item $\gamma=-2.4_{-5.4}^{+3.7}$.
\end{itemize}

\begin{figure}[h]
\includegraphics[width=1.0\linewidth]{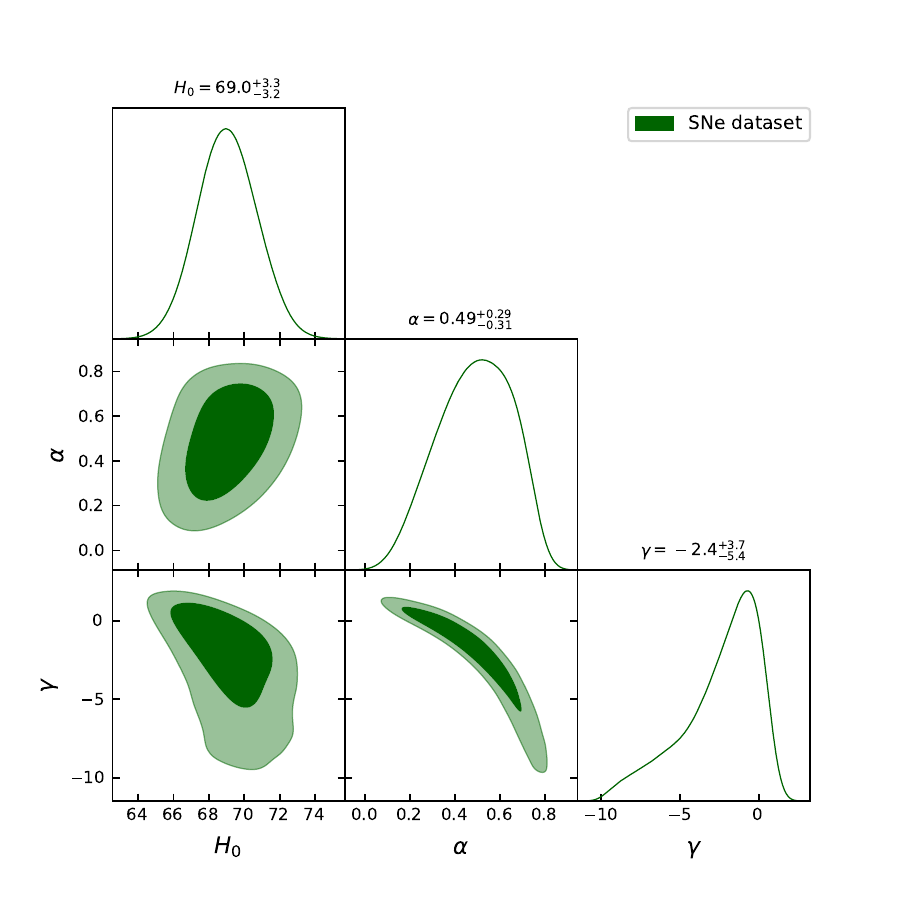}
     \caption{Contours of likelihood for $SNe$ datasets, indicating $1-\sigma$ and $2-\sigma$ confidence regions.}\label{F_SNe}
\end{figure}

And, for the combined analysis encompassing $Hz$, $SNe$, and $BAO$ dataset:
\begin{itemize}
    \item $H_0=68.0_{-1.4}^{+1.5}$,
    \item $\alpha=0.310_{-0.041}^{+0.043}$,
    \item $\gamma=-0.02_{-0.46}^{+0.45}$.
\end{itemize}

\begin{figure}[h]
\includegraphics[width=1.0\linewidth]{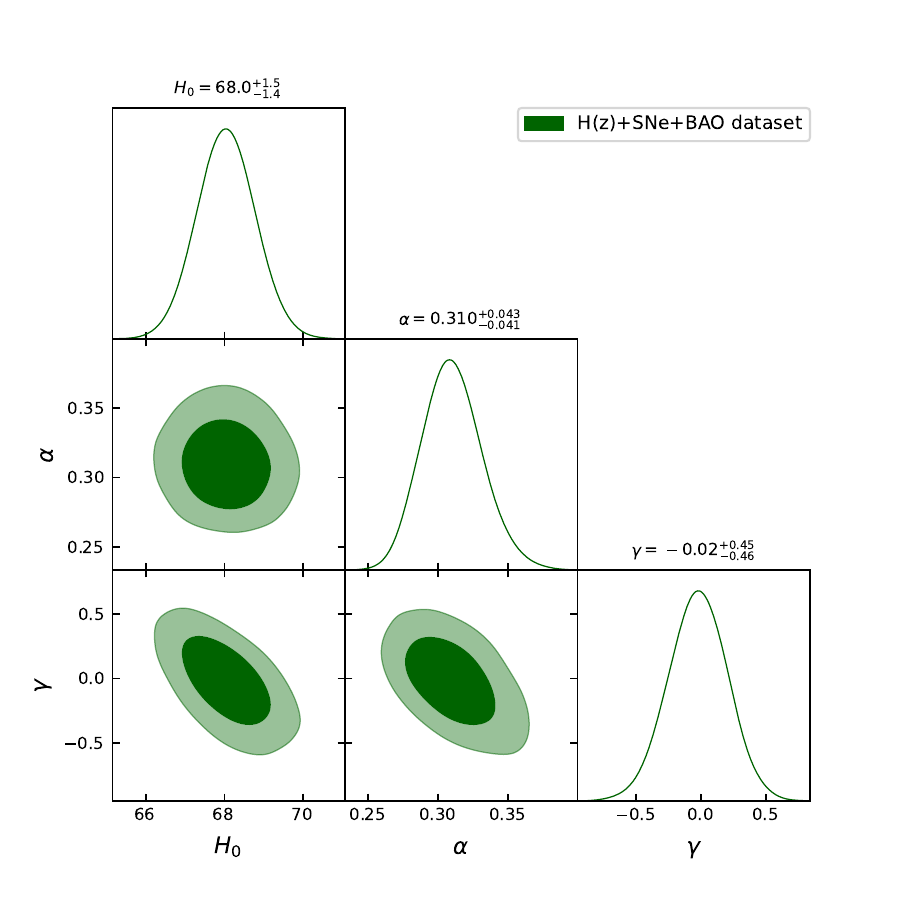}
     \caption{Contours of likelihood for $Hz+SNe+BAO$ datasets, indicating $1-\sigma$ and $2-\sigma$ confidence regions.}\label{F_combine}
\end{figure}

Figs. \ref{F_Hz}-\ref{F_combine}, derived from the MCMC analysis, depict the error-encapsulated likelihood contours at the $1$-$\sigma$ and $2$-$\sigma$ confidence levels for these parameter values. We observe that the constrained values for $H_0$ and $\alpha$ are relatively consistent across the different datasets. However, the values of $\gamma$ exhibit significant variation among the three datasets. Specifically, for the $Hz$ data, $\gamma$ is approximately $-0.2$, while for the $SNe$ data, it is approximately $-2.4$. In contrast, for the combined $Hz+SNe+BAO$ dataset, the value of $\gamma$ is around $-0.02$. These substantial differences in the values of $\gamma$ across the datasets are noteworthy as they suggest potential deviations from the $\Lambda$CDM model. 

\begin{figure}[h]
\includegraphics[scale=0.7]{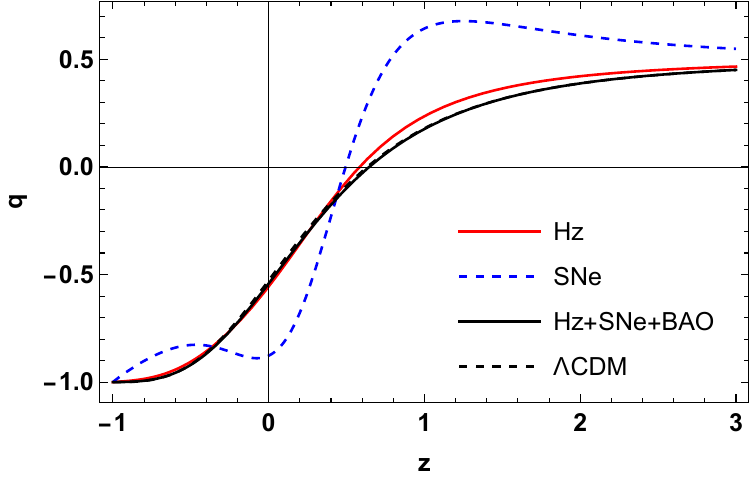}
\caption{This figure illustrates the evolutionary behavior of the deceleration parameter $q(z)$ as a function of redshift $z$, based on constrained values derived from the datasets of $Hz$, $SNe$, and $Hz+SNe+BAO$ datasets.}\label{F_q}
\end{figure}

Furthermore, the cosmological observations have revealed a crucial phenomenon: the acceleration of cosmic expansion. When DE is either absent or has minimal influence, a cosmological model must encompass a deceleration phase during the early matter-dominated era. This deceleration phase is crucial as it facilitates the gravitational aggregation of matter, which is fundamental for the universe's structure formation. Consequently, a comprehensive cosmological model must account for both deceleration and accelerated expansion phases to provide a holistic account of the universe's developmental history. In this context, the deceleration parameter assumes a central role. In Fig. \ref{F_q}, we present the deceleration parameter plot against the redshift within our model for constrained model parameter values derived from the analyses of the $Hz$, $SNe$, and the combined $Hz$, $SNe$, and $BAO$ datasets. This representation vividly illustrates the universe's transition from a decelerating phase to an accelerating one. For the $SNe$ dataset, it is observed that around $z=0$, there is a phase where the acceleration rate is reduced for a period before the curve sharply increases and then declines. This behavior could be interpreted in the context of DE. The observed pattern might suggest a transition in the properties of DE at low redshifts, resulting in a temporary deviation from the expected acceleration pattern. Furthermore, it offers insights into the present values of the deceleration parameter for three distinct observational datasets: $Hz$, $SNe$, and the combined analysis of $Hz$, $SNe$, and $BAO$. These values, along with their associated errors, are as follows: $q_0=-0.56_{-0.56}^{+0.53}$, $q_0=-0.88_{-0.38}^{+0.57}$, and $q_0=-0.54_{-0.21}^{+0.21}$, respectively. Remarkably, these findings are in alignment with results previously reported in the scientific literature \cite{Garza,Cunha,Camarena}, providing robust support for our understanding of the universe's expansion dynamics. In addition, the transition redshift ($z_{tr}$), which occurs during a phase transition and is primarily influenced by variations in model parameters, is explored. For the parameters constrained by the $Hz$, $SNe$, and combined $Hz+SNe+BAO$ datasets, the transition redshift values are as follows: $z_{tr}=0.59_{-0.3}^{+0.3}$, $z_{tr}=0.49_{-0.07}^{+0.07}$, and $z_{tr}=0.67_{-0.07}^{+0.08}$. These outcomes are in harmony with values previously reported in the scientific literature \cite{Mehrabi}, further solidifying the robustness of our cosmological framework and its alignment with established cosmological paradigms.

\begin{figure}[h]
\includegraphics[scale=0.7]{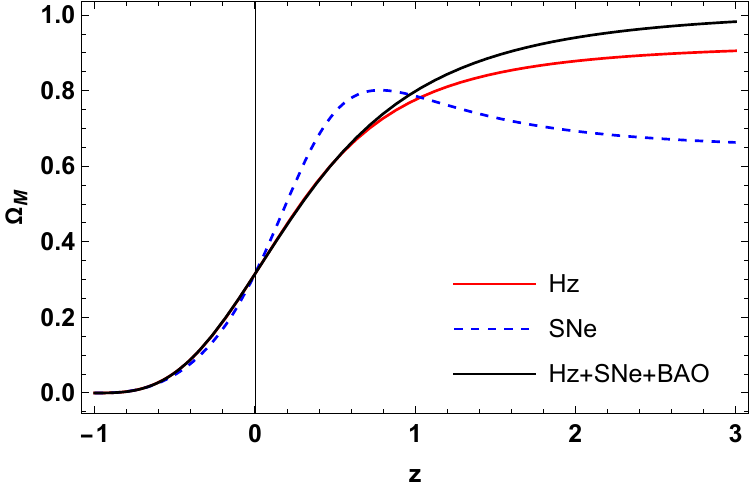}
\caption{This figure illustrates the evolutionary behavior of the density parameter for matter $\Omega_{M}(z)$ as a function of redshift $z$, based on constrained values derived from the datasets of $Hz$, $SNe$, and $Hz+SNe+BAO$ datasets.}\label{F_Omegam}
\end{figure}

\begin{figure}[h]
\includegraphics[scale=0.7]{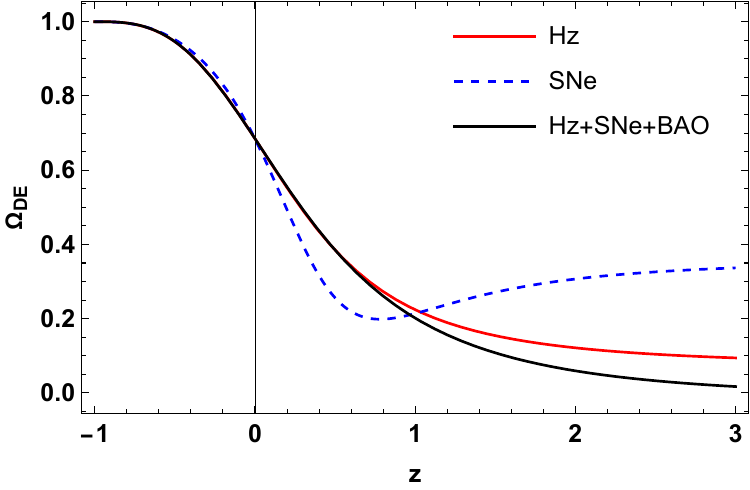}
\caption{This figure illustrates the evolutionary behavior of the density parameter for DE $\Omega_{DE}(z)$ as a function of redshift $z$, based on constrained values derived from the datasets of $Hz$, $SNe$, and $Hz+SNe+BAO$ datasets.}\label{F_OmegaDE}
\end{figure}

Figs. \ref{F_Omegam} and \ref{F_OmegaDE} provide a visual representation of the temporal evolution of the density parameters for matter and DE, respectively. These graphical representations offer valuable insights into the composition of the Universe. At the outset, the Universe is predominantly characterized by the dominance of non-relativistic matter, encompassing both DM and baryonic matter. During this phase, the contribution from the DE density parameter is minimal, almost negligible. However, as the Universe undergoes expansion, the density parameter associated with matter gradually diminishes, influenced by the Universe's increasing size. Over time, a significant transformation occurs. The density parameter attributed to DE steadily gains prominence and begins to outweigh the contribution from matter. This pivotal shift in dominance marks a critical juncture in cosmic evolution, ultimately propelling the Universe into a phase of accelerated expansion.

\begin{figure}[h]
\includegraphics[scale=0.7]{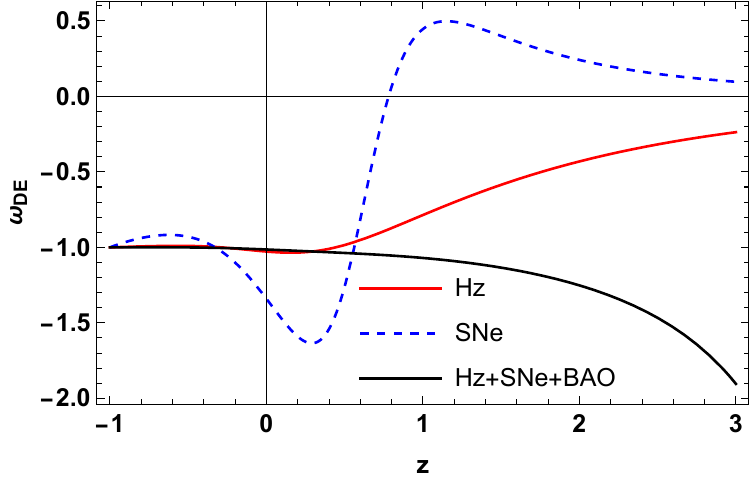}
\caption{This figure illustrates the evolutionary behavior of the EoS parameter for DE $\omega_{DE}(z)$ as a function of redshift $z$, based on constrained values derived from the datasets of $Hz$, $SNe$, and $Hz+SNe+BAO$ datasets.}\label{F_EoS}
\end{figure}

In addition to the insights provided earlier, we delve further into our analysis by examining the EoS parameter for DE, denoted as $\omega_{DE}=\frac{p_{DE}}{\rho_{DE}}$, which is illustrated in Fig. \ref{F_EoS}. It's worth noting that this parameter plays a critical role in our understanding of the universe's dynamics. As a reminder, the condition ($\omega < -\frac{1}{3}$) is of paramount importance, as it encompasses both the quintessence regime ($-1<\omega_{DE} < -\frac{1}{3}$) and the more exotic phantom regime ($\omega_{DE} < -1$). This condition serves as the fundamental criterion for identifying an accelerating universe, a phenomenon that has profound implications for our cosmological framework. Furthermore, the EoS parameter $\omega_{DE}$ also provides crucial insights into the nature of vacuum energy or the cosmological constant. In cases where $\omega_{DE}=-1$, it characterizes the vacuum energy or the cosmological constant, offering a glimpse into the mysterious and dominant force responsible for driving the universe's accelerated expansion. This aspect of our analysis brings us closer to unraveling the enigmatic nature of the universe's DE and its influence on cosmic evolution. Through the process of model fitting to observational data (including measurements of $Hz$, $SNe$, and the combined $Hz+SNe+BAO$), we have determined the present value of the EoS parameter for different sets of constrained model parameters. These values are as follows: $\omega_{DE}^{0}=-1.03_{-0.54}^{+0.51}$, $\omega_{DE}^{0}=-1.34_{-1.34}^{+1.16}$, and $\omega_{DE}^{0}=-1.01_{-0.2}^{+0.2}$, respectively \cite{Mukherjee1,Novosyadlyj/2012,Suresh/2014}.

In Fig. \ref{F_EoS}, we depict the evolution of $\omega_{DE}$, shedding light on intriguing aspects of DE behavior. Notably, our analysis reveals that $\omega_{DE}$ transitions from the quintessence region ($\omega_{DE}>-1$) to the phantom regime ($\omega_{DE}<-1$) within the redshift range of -0.6 to 0.2. This transition has been confirmed for both $Hz$ and $SNe$ data, illustrating a shift from quintessence to phantom behavior \cite{Wu}. However, when considering the combined dataset, the EoS parameter $\omega_{DE}$ consistently resides in the phantom regime throughout. At lower redshifts, this evolution of the EoS parameter aligns favorably with various observational datasets \cite{Melchiorri/2003,Alam/2004}. Furthermore, the cosmological dynamics of a universe featuring phantom energy present many captivating features \cite{Caldwell/2003}. In-depth analyses of Lagrangians describing phantom energy have indicated that, depending on specific conditions, such a universe can either culminate in a dramatic 'big rip' or asymptotically approach a de Sitter expansion \cite{McInnes}. Importantly, these outcomes are not universal but rather contingent on the parametrization of the Hubble parameter. Refs. \cite{Vagnozzi/2020,Valentino/2016} suggest that a phantom-like component with an effective EoS $\omega=-1.29$ has the potential to resolve current tensions between the Planck data set and other priors within an extended $\Lambda$CDM scenario. Notably, our model's estimated range for $\omega_{DE}$ falls within the phantom phase, spanning from $-1.01$ to $-1.34$. Consequently, our model may hold promise in alleviating some of the present tensions in the cosmological landscape.

\section{Final Remarks}
\label{sec6}

The exploration of cosmic evolution through the reconstruction approach has been a focal point in cosmological research. Researchers have rigorously investigated two distinct methods: parametric and non-parametric reconstruction. At present, no single gravity theory has emerged as a definitive explanation for the myriad phenomena observed in the universe. Consequently, both of these reconstruction approaches offer valuable perspectives, each with its own merits in addressing this complex cosmic puzzle. 

Within the context of our study, we have introduced a cosmological model of the FLRW universe, employing the parametric approach. Parametric methods have demonstrated notable success in elucidating the universe's evolutionary trajectory, spanning from its early deceleration to its subsequent acceleration. This track record positions parametrization as a promising avenue for effectively comprehending and constructing future cosmological scenarios. Our primary objective in this research endeavor is to undertake the reconstruction of the Hubble parameter. This endeavor enables us to meticulously trace the ongoing evolution of the present-day universe. To accomplish this, we work within the framework of a spatially flat, homogeneous, and isotropic FLRW spacetime, which serves as the backdrop for our cosmological exploration.

In Sec. \ref{sec3}, our approach involved using a parametrization technique to reconstruct the Hubble parameter. Importantly, this parametrization method is not influenced by any preconceived assumptions regarding the nature of DE. Within this parametric framework, the $\Lambda$CDM model represents a specific instance characterized by particular parameter values: $\alpha=\Omega_{M0}$, $\beta=\Omega_{\Lambda}$, and $\gamma=0$. Then, we subjected the Hubble parameter to rigorous scrutiny by comparing it with the most up-to-date observational datasets, specifically the $Hz$, $SNe$, and the combined $Hz+SNe+BAO$ datasets. Our aim was to place constraints on the model parameters $H_0$, $\alpha$, and $\gamma$. The outcomes of this analysis are depicted in Figs. \ref{F_Hz} to \ref{F_combine}, which illustrate the best-fit values of the model parameters as well as the $1-\sigma$ and $2-\sigma$ confidence regions. Throughout the cosmic evolution, the deceleration parameter undergoes a significant transition from deceleration to acceleration. Notably, our analysis for the combined $Hz+SNe+BAO$ datasets yielded a deceleration parameter of $q_0=-0.54_{-0.21}^{+0.21}$, a value that closely resembles the prediction of the $\Lambda$CDM model. Furthermore, it's worth highlighting that our model is situated within a phantom phase, characterized by an EoS spanning from $-1.01$ to $-1.34$ (see Fig. \ref{F_EoS}). This transition from the quintessence region to the phantom regime is evident when analyzing both $Hz$ and $SNe$ datasets. However, when we take into account the combined dataset, an interesting pattern emerges. The EoS parameter consistently remains within the phantom regime, showing a robust propensity toward phantom behavior throughout the analysis. It becomes evident that our constructed model accommodates a broad spectrum of $\omega_{DE}$ values. As referenced in \cite{Vagnozzi/2020,Valentino/2016}, it has been suggested that introducing a phantom-like component with an effective EoS $\omega=-1.29$ may serve as a potential resolution for the existing tension between the Planck data set and other prior observations within an extended $\Lambda$CDM framework.

Finally, we have noted that our reconstructed model for the Hubble parameter falls within the phantom phase, characterized by an EoS spanning from $-1.01$ to $-1.34$. Consequently, our model may hold the potential to mitigate some of the existing tensions in the current cosmological framework. The disparities in both the values and behaviors of the EoS parameter compared to the $\Lambda$CDM model indicate the emergence of a novel DE alternative. This reconstruction of the Hubble parameter offers a promising avenue for explaining the accelerated expansion of the universe and provides new insights into cosmic phenomena.

\section*{Acknowledgments}
This work was supported and funded by the Deanship of Scientific Research at Imam Mohammad Ibn Saud Islamic University (IMSIU) (grant number IMSIU-RPP2023082).
\section*{Data availability} All data used in this study are cited in the references and were obtained from publicly available sources.

\end{document}